\documentstyle[preprint,aps]{revtex}

\begin{document}
\bibliographystyle{prsty}
\draft

\title{\bf Quantum computing with four-particle decoherence-free states in 
ion trap} 
\author{Mang Feng$^{1,2}$
\thanks{Electronic address: feng@isiosf.isi.it},
Xiaoguang Wang$^{2}$
\thanks{Electronic address: xgwang@isiosf.isi.it}}
\address{$^{1}$ Laboratory of Magnetic Resonance and Atomic and Molecular
Physics,\\  Wuhan Institute of Physics and Mathematics, Academia Sinica,
Wuhan, 430071, P.R.China \\
$^{2}$ Institute for Scientific Interchange Foundation, \\ Villa Gualino, Viale Settimio 
Severo 65, I-10131, Torino, Italy }
\date{\today}
\maketitle

\begin{abstract}

Quantum computing gates are proposed to apply on trapped ions in
decoherence-free states. As phase changes due to time evolution of components
with different eigenenergies of quantum superposition are completely frozen,
quantum computing based on this model would be perfect. Possible application of our 
scheme in future ion-trap quantum computer is discussed.

\end{abstract} 
\vskip0.1cm 


Much effort has been made on quantum computing over past few years due to
great advantages of quantum computing over the computing made in
existing computers for solving classically intractable problems [1,2] and
finding tractable  solutions more rapidly [3]. Some system [4], such as nuclear
magnetic resonance (NMR), trapped ions, cavity quantum
electrodynamics (QED) and optical photons have been proven to meet the
requirement of quantum computing.

There are some special advantages respectively for each systems referred to above, while we
will focus in the present work on the discussion of quantum gates on trapped ions. Since 
the proposal of Cirac and Zoller [5], many elaborate ideas have been put forward on quantum
computing with trapped ions [6,7]. There are some obvious advantages in using trapped ions to make 
quantum computing: first, the manipulation can be made on individual ions by quantum level [8], 
rather than in NMR the operation performed on the bulk molecules; second, the 
entanglement between ions is deterministic [9,10], which can avoid the problem due to 
the selection of data from random process in experiments with photons; third, 
there is effective interaction in the trapped ion system, which can avoid the problem of 
incomplete measurement of Bell states due to the lack of photon-photon interaction [11,12]; 
fourth, the strict requirement for ion-trap quantum computing, i.e., cooling down trapped 
ions to the ground vibrational state, has been loosened. The so-called hot-ion quantum 
computing, in which internal states of ions acting as qubits are completely decoupled 
from vibrational states of ions, has been realized experimentally at the scale of four 
ions [10]. Subsequent experiments [13,14] for testing the Bell inequality and constructing the 
decoherence-free qubits are also based on hot-ion quantum computing model. 

In the present work, we will try to perform quantum computing gates on ion-pairs encoded in decoherence-free 
states, as referred to in Ref.[14]. In a preliminary work [7], the effective quantum computing gates 
have been performed on two identical two-level ions both illuminated with two lasers of different 
frequencies $\omega_{1,2}=\omega_{0}\pm\delta$, where $\omega_{0}$ is the resonant transition 
frequency of the ions, and $\delta$ the detuning, not far from the trap frequency $\nu$. With the 
choice of laser detunings, the only energy conserving transition is from $|ggn>$ to $|een>$ 
or from $|gen>$ to $|egn>$, where the first (second) letter denotes the 
internal state $e$ or $g$ of the $i$th ($j$th) ion and $n$ is the quantum 
number for the vibrational state of the ion. As we consider $\nu-\delta\gg\eta\Omega$ with $\eta$ 
being the Lamb-Dicke parameter and $\Omega$ the Rabi frequency denoting the interaction of ions 
with lasers, there is only negligible population being transferred to the intermediate states with 
vibrational quantum number $n\pm 1$. Therefore this two-photon process has nothing to do with the 
vibrational state $|n>$ and quantum computing with such configuration is valid even in the case that 
ions are in thermal states. Based on that model, the decoherence-free states $|eg>\pm e^{i\theta}|ge>$
with $\theta$ arbitrary phase were experimentally tested, which showed that qubits encoded in 
decoherence-free states is robust against the environment-induced dephasing [14].
The hot-ion quantum computing model can be generalized to the case of 
large numbers of ions. As referred to above, in NIST group, the entangled state of four ions with 
the form of $\frac {1}{\sqrt{2}}(|eeee>+i|gggg>)$ has been experimentally realized [10]. In fact, by setting 
different initial conditions, we would obtain other entangled states with decoherence-free forms 
of $\frac {1}{\sqrt{2}}(|egeg>+i|gege>)$ and $\frac {1}{\sqrt{2}}(|egge>+i|geeg>)$, and so on. 
In what follows, we will show how to achieve basic gates of quantum computing on those states and discuss how
to use our scheme in an actual ion-trap quantum computing.

The effective Rabi frequency for transitions between product states of four-ion string can 
be calculated by means of second order perturbation formula, 
$$\tilde{\Omega} = 2[\frac {<egeg~n|H_{int}|gggg~n+2><gggg~n+2|H_{int}|gege~n>}{-(2\nu-\delta)} +$$
\begin{equation}
\frac {<egeg~n|H_{int}|eeee~n-2><eeee~n-2|H_{int}|gege~n>}{(2\nu-\delta)}]=(2n+1)
\frac {(\Omega\eta)^{2}}{2\nu-\delta}.
\end{equation}
where 
\begin{equation}
H_{int}=\frac {\Omega}{2}\sum_{j\neq k,
j,k=1}^{4}(\sigma_{j}\sigma_{k}e^{-i[\eta (a^{2}+a^{+2})-\omega_{L}t]}+H.c)
\end{equation}
with $\Omega$ being usual Rabi frequency of laser-ion interaction, $\eta$ the Lamb-Dicke parameter 
and $\omega_{L}$ the laser frequency. For the transition between $|egge>$ and $|geeg>$, we can obtain 
the same value of $\tilde{\Omega}$ as in Eq.(1). It is obvious from Eq.(1) that the transitions between 
different four-particle states are dependent on the vibrational state of ions. So to some extent, it is 
not a true hot-ion quantum computing. However, as $n$ keeps unchanged at the beginning and at the end of the
process, as long as $\Omega\eta\ll (2\nu-\delta)$, the hot-ion 
quantum computing can still be carried out in the case that $n$ is not very large. If $n=0$, the effective 
Rabi frequency of four-particle case is almost half that of 
two-particle case. Moreover, straightforward deduction from Eq.(2) can yield the time evolution of the 
four-particle case, similar to those shown in Refs.[7,10],
$$|egeg>\Longrightarrow \cos(\frac {\tilde{\Omega t}}{2})|egeg>-i\sin(\frac {\tilde{\Omega t}}{2})|gege>;$$
$$|gege>\Longrightarrow \cos(\frac {\tilde{\Omega t}}{2})|gege>-i\sin(\frac {\tilde{\Omega t}}{2})|egeg>;$$
$$|egge>\Longrightarrow \cos(\frac {\tilde{\Omega t}}{2})|egge>-i\sin(\frac {\tilde{\Omega t}}{2})|geeg>;$$
\begin{equation}
|geeg>\Longrightarrow \cos(\frac {\tilde{\Omega t}}{2})|geeg>-i\sin(\frac {\tilde{\Omega t}}{2})|egge>.
\end{equation}

To show our scheme is available to implement quantum computing, we have to demonstrate how to achieve 
controlled-NOT (CN) gate and Hadamard one [15]. As it is relatively easy to implement
a Hadamard gate, the main work in this respect is how to achieve a CN gate.  
Similar to Ref.[7] that the CN gate is performed on two identical trapped ions experiencing
a single-mode laser beam, we can also carry out the CN gate on the four identical ions based on Eq.(3). If 
the computational space is formed by four states $|egeg>_{1234}$, $|gege>_{1234}$, $|egge>_{1234}$ and 
$|geeg>_{1234}$ where subscripts mean the labeling of ions, with following sequence of
operations: $H_{34}, P_{34}, R, P_{34}, H_{12}, H_{34}$ and $P^{-1}_{34}$, we will find the CN gate as follows 
$$|eg>_{12}|ge>_{34}\rightarrow |eg>_{12}|ge>_{34},$$ 
$$|ge>_{12}|ge>_{34}\rightarrow |ge>_{12}|ge>_{34},$$ 
$$|eg>_{12}|eg>_{34}\rightarrow |ge>_{12}|eg>_{34}$$ 
and
$$|ge>_{12}|eg>_{34}\rightarrow |eg>_{12}|eg>_{34}$$
where $H_{ij}$ is the transformations of $|eg>_{ij}\rightarrow (|eg>_{ij}-i|ge>_{ij})/\sqrt{2}$ and 
$|ge>_{ij}\rightarrow (|ge>_{ij}-i|eg>_{ij})/\sqrt{2}$ made in Ref. [10] by radiating the two ions $i$ and $j$ 
simultaneously. $P_{ij}$ is a $\frac {\pi}{2}$ phase change of 
$|eg>_{ij}$ and $R$ is the evolution shown in Eqs.(3) with $\Omega t=3\pi/4$.

The Hadamard gate on a ion-pair can be easily realized by $H_{ij}$ in addition to $P_{ij}$, that is,
$|eg> \rightarrow i(|eg>-|ge>)/\sqrt{2}$ and $|ge> \rightarrow (|ge>|+|eg>)/\sqrt{2}$. Therefore
any quantum computing operation can be constructed with our scheme.
However, the main problem of our scheme is that the CN gate is restricted to be performed on two neighboring
ion-pairs, which is also the common drawback of various hot-ion quantum computing schemes.
It is recalled that  in Cirac-Zoller scheme and other similar models, the center-of-mass (COM) state of 
trapped ions plays the role of data bus in performing the CN gate on two separate qubits. In contrast 
the vibrational states of ions are excluded from the computational space in our scheme.
 Although we can implement Grover search by using such a CN gate, like in Ref.[16], it is 
impossible for us to accomplish Shor algorithm because Shor algorithm includes quantum Fourier transform, 
and to perform quantum Fourier transform we have to carry out a series of Hadamard gates and CN ones, where 
most CN gates are not performed on neighboring qubits. To solve this problem, we must transfer the states
of qubits by swapping operation. But we can do that more directly, as explained later.

It is still a great challenge both theoretically and experimentally to scale up the technique suitable for a 
few ions to a large-scale quantum computer with thousands of ions [17]. The computing operation on ions would
become more and more complicated with the increase of the number of ions [18,19]. Moreover, with the increase
of the number of ions, the spatial separation of trapped ions decreases, which makes their spatial resolution
more and more difficult [20]. So to solve the problem, we have to use multiple traps. One of the promising 
schemes in this respect was proposed by NIST group [21], in which each ion-trap quantum computer consists of 
two regions. One of them is called storage region, which stores thousands of ions
and acts as quantum register. The other is called accumulator region, in which quantum logic operations take
place. As we only work with a few qubits at a time in accumulator region, we can avoid the slow gate
speeds and detection inefficiency. Suppose that we implement our ion-pair scheme in such a quantum computer.
After a piece of time for quantum computing, we hope to transfer the message (i.e. the state representing
certain result of quantum computing we have made) of a ion-pair (for example ions 1 and 2) in accumulator 
region to another ion-pair (ions 3 and 4) in storage
region, so that ions 1 and 2 can be reset to new initial states and used for the coming computation. We can
use teleportation [22] for doing that job. To this end, we prepare many four-ion entangled states 
$\frac {1}{\sqrt{2}}(|egge>-i|geeg>)$ in storage region for the thousands of trapped ions. Suppose state of 
ions 1 and 2 to be $\Psi=|eg>+e^{i\theta}|ge>$ [23], and four ions 3,4,5 and 6 to be entangled as 
$\frac {1}{\sqrt{2}}(|egge>-i|geeg>)$ in storage region. By setting 
$|\tilde{1}>=|eg>$, $|\tilde{0}>=|ge>$, and Bell states of 
$|\tilde{1}\tilde{1}>\pm i|\tilde{0}\tilde{0}>$ and $|\tilde{1}\tilde{0}>\pm i|\tilde{0}\tilde{1}>$,
 moving adiabatically ions 5 
and 6 to accumulator region will transfer the message from ions 1 and 2 to ions 3 and 4 by means of the
idea of teleportation. Of course with the transfer of states between different ion-pairs, the effect of CN 
gate can be on two separate ion-pairs. 
As both quantum computing and teleportation can be made without any affect from dephasing, our scheme 
would be the first step to the building of an actual ion-trap quantum computer with more qubits.  

Another problem is that the number of ions used for quantum computing in our scheme is twice that in former
schemes, which is also a challenge to the present ion trap technique unavailable to confine large numbers of 
ions. Besides, we also note that, although the decoherence-free states keep decoherence away from our desired
 states theoretically, some unclear source of decoherence existing in the present experiments would
affect our scheme. As referred to in Ref.[10], even if both the center-of-mass and stretch states are cooled 
to ground states, this kind of decoherence affects the four-ion experiment more strongly than the two-ion one. 
It is speculated that this kind of decoherence is mainly resulted from laser fluctuations. But in the present
scheme we speculate that the four-ion states transition related to
the vibrational state would be probably another source of decoherence. Although mathematically we can neglect
the vibrational state as it remains unchanged at the beginning and at the end of our scheme, the influences 
from the vibrational state do 
exist due to the incompletely destructive interference between the paths of $\pm(2\nu-\delta)$. Therefore, 
for experimentally achieving our scheme, more advanced technique is highly demanded to remove the laser 
fluctuation and keep the vibrational state coherent for a longer time. 
    
Nevertheless, compared with former various schemes, the distinct character of our scheme is that 
quantum computing would be made perfectly, without any possible phase change and to the largest extent 
excluding the detrimental effect of decoherence. More importantly, unlike both the incomplete measurement 
of Bell states due to the lack of effective photon-photon interaction and partial distinguishablility of Bell
states in cavity-atom system [24], we can make a complete measurement of Bell states in teleportation with 
our scheme. From
Eq.(3) we know, a laser pulse on ions 1, 2, 5 and 6 with the period of $t=\pi/2\tilde{\Omega}$ would yield 
$(|egeg> + i|gege>) \rightarrow |egeg>$, $(|egeg> - i|gege>)\rightarrow -i|gege>$, 
$(|geeg> + i|egge>)\rightarrow |geeg>$ and $(|geeg> - i|egge>)\rightarrow i|egge>$. So by 
detecting internal levels of ions 1, 2, 5 and 6 respectively, we know that, if the result is $|egeg>$, the  
ions 3 and 4 has been in the desired state of $\Psi$. But if results are respectively, $|gege>$, $|geeg>$ or 
$|egge>$, we have to perform operations $\sigma_{z}^{3}$, $\sigma_{x}^{3}\sigma_{x}^{4}$, and 
$\sigma_{x}^{3}\sigma_{x}^{4}\sigma_{z}^{3}$ respectively on ions 3 
and 4 to get the state of $\Psi$. As this teleportation is made with 100$\%$ success probability, 
the message can be transferred safely and completely from accumulator region to storage one.

Let us have a brief discussion for the technique respect of our scheme. To perform a CN gate, we need time 
at least $t= \frac {3\pi}{2}\frac {2\nu-\delta}{(\Omega\eta)^{2}}$. As $\nu\approx \delta$, 
$\eta=0.23/N^{2} (N=4$ in our work), and $\Omega=2\pi\times 500 KHz\approx 0.1\nu$ [10,14], we have 
$t=7\times 10^{-4} s$, which is much shorter than the lifetime ( 1$s$ to 10 $s$) of the metastable level of 
a trapped ion which acts as the excited 
state of a qubit. So our CN gate will work very well. But if we want to transfer the message from accumulator
region to storage region, we must take time to move ion-pairs adiabatically from storage region to accumulator
region and cool the moved ion-pairs after they get to the accumulator region. This time should not be larger 
than $1\sim 10$ $s$. However, how to effectively move trapped ions from one place to another place with minimum 
heating is still a open question.

In summary, a robust scheme for performing a perfect quantum computing has been proposed in trapped ions
system. The entanglement of four trapped ions in our scheme is of the different form from that
achieved by NIST group, but our entanglement is more important because it corresponds to
decoherence-free states. As the vibrational state is only virtually excited although 
transitions between different four-ion states are formally dependent on the vibrational state of ions, it is 
not strictly required to cool the ions to vibrational ground states for achieving our scheme, whereas  
quantum number of vibrational state of ions can not be too large. For avoiding any possible
decoherence, we had better to cool the ions to the vibrational state closed to $n=0$ before implementing our 
scheme. Moreover, we benefit from decoherence-free states not only in the phase stability during the period 
of quantum computing and teleportation, but also in the safe storage of information. 

The work is partly supported by the National Natural Science Foundation of China.


\begin{thebibliography}{27}
\bibitem{23} Here we consider the simplest case that at the moment we want to transfer states of qubits in
accumulator region, the states of those qubits are of product form. If some of the states are entangled
states, the tranfer of states can still work by repeatly using teleportation between different ion-pairs
in accumulator region and storage one.
\end{thebibliography}
\end{document}